\documentclass[peerreview,onecolumn]{IEEEtran}

\usepackage{cite}      

\usepackage{graphicx}
\usepackage{psfrag}
\usepackage{amsmath}
\usepackage{amscd,amssymb}   
\usepackage{setspace}
\begin{document}
%
\title{Laser Doppler Velocimetry for Joint Measurements of Acoustic and Mean Flow Velocities : LMS-based Algorithm and CRB Calculation}
%
%

\author{Laurent Simon,~\IEEEmembership{Member,~IEEE,}
         Olivier Richoux,
        Anne Degroot and Louis Lionet}

\maketitle

\begin{abstract}

This paper presents a least mean square (LMS) algorithm for the joint estimation of acoustic and mean flow velocities from laser doppler velocimetry (LDV) measurements. The usual algorithms used for measuring with LDV purely acoustic velocity or mean flow velocity may not be used when the acoustic field is disturbed by a mean flow component. The LMS-based  algorithm allows accurate estimations of both acoustic and mean flow velocities. The Cram\'er-Rao bound (CRB) of the associated problem is determined. The variance of the estimators of both acoustic and mean flow velocities is also given. Simulation results of this algorithm are compared with the CRB and the comparison leads to validate this estimator.

\end{abstract}

\begin{keywords}
Laser Doppler Velocimetry, Acoustic Velocity, Mean Flow Velocity, Least Mean Square Method, Cram\'er-Rao Bound 
\end{keywords}

\IEEEpeerreviewmaketitle

\section{Introduction}

Laser Doppler Velocimeter (LDV) is an optical technique allowing direct measurement of local and instantaneous fluid velocity. This method is nonintrusive and is based on optical interferometry for estimating the velocity of scatterers suspended in a fluid by means of the frequency analysis of the light scattered by the seeding particles \cite{Albrecht}.

For fluid mechanics measurements, the particle velocity can be considered as constant during the transit time of the seeding particle through the measurement volume (defined by the interferometry fringes volume) and the frequency of the LDV signal is constant during this period \cite{Buchhave}. Typical order of magnitude of mean flow velocities are from a few meters per second up to higher than the acoustic celerity (supersonic flow). The data processing consists then to estimate the power spectral density (PSD) of the velocity signal, from Poisson-based randomly distributed samples. PSD may be estimated by interpolating the randomly distributed samples, by resampling the interpolating signal and by compensating the effect of interpolation in the Fourier domain \cite{Boyer}-\cite{Simon}. The autocorrelation function (ACF) may also be reconstructed from the randomly distributed samples and the Fourier transform of the estimated ACF gives an estimation of the PSD \cite{Nobach}. Lastly, Kalman filtering may be used for estimating the PSD \cite{Banning}.

For sine acoustic excitation, the particle velocity is no longer constant and the LDV signal is frequency modulated \cite{Taylor76}-\cite{Greated}. To estimate the particle velocity from these signals, specific signal processing techniques are used as spectral analysis \cite{Taylor81,Davis,Vignola}, photon correlation \cite{Sharped89} or frequency demodulation associated to post-processing methods \cite{Valiere00,Valeau04,Gazengel05}. Typical order of magnitude of mean flow velocities are from a few micrometers per second up to $100$ millimeters per second, for frequencies in $[10-4000]$ Hz.

On one hand, for most acoustic measurements, the particle velocity can be considered as the sum of an AC-component due to acoustic excitation and a weak DC-contribution due to flow. When the particle oscillates in the measurement volume during further acoustic periods, the effect of the flow can be reduced and usual post-processing methods may be used \cite{Valiere00}-\cite{Gazengel03}.

On the other hand, the DC-flow component prevents in many cases the use of the post-processing methods given by \cite{Valiere00,Valeau04,Gazengel05}, because the signal time length is less or largely less than one acoustic period. The aim of this paper is to estimate both the dc (flow) and ac (acoustic) components from such LDV signals.

Lazreq and Ville \cite{Lazreq} measured the acoustic velocity in presence of mean flow by means of a probe consisting in a hot wire and a microphone. Their results showed a good agreement between the theory and the experiment but this probe cannot be considered as nonintrusive. LDV has also been used by adapting the slotting technique to estimate the acoustic particle velocity in a turbulent flow \cite{Minotti} with a 2D-LDV velocimeter. The acoustic impedance was estimated by means of a LDV probe and with a microphone probe and the different results were compared. Finally, Boucheron \textit{et al} \cite{Boucheron} has developed a new method of signal processing called 'perio-correlation' in order to estimate sine acoustic velocity in strong mean flow  by LDV.

In this work, the sine acoustic excitation is supposed to be perfectly known and a frequency demodulation technique \cite{Gazengel05} is performed to estimate the particle velocity from the LDV signal. In this paper, we propose a new method to estimate jointly the acoustic particle velocity (amplitude and phase) and the mean flow velocity from the velocity signal. This method is based on the least mean square (LMS) algorithm. The mean flow velocity, the amplitude and phase of acoustic particle velocity are estimated for each seeding particle crossing the measurement volume. Furthermore, the Cram\'er-Rao bound (CRB) of the associated problem is calculated. The CRB gives the lowest variance of any unbiased estimator and consequently yields theoretically the minimum uncertainties linked to the velocity estimations (acoustic and mean flow velocities). Lastly, simulated data are processed, in order to validate the LMS-based algorithm and to compare the variance of the results with the Cram\'er-Rao bound.

Section II deals with the LDV principles including the velocity signal modeling and the associated signal processing for acoustic applications. In section III, the data processing based on the least mean square algorithm is explained and the Cram\'er-Rao bound of both the mean flow and acoustic velocities are determined. Finally, the results of the Monte Carlo simulation are shown and compared to the Cram\'er-Rao bounds in section IV, for acoustic frequencies in $[125-4000]$ Hz, for acoustic velocities in $[0.05-50]$ mm.s$^{-1}$ and for mean flow velocities in $[0.05-5000]$ mm.s$^{-1}$.

\section{Fundamentals of Laser Doppler Velocimetry}
\label{section1}

In this section, we consider time-varying signals such that $t \in [t_q - T_q/2,t_q + T_q/2]$, $t_q$ being the central time of the signal, $T_q$ being a time of flight, and $q$ being associated to a given seeding particle. 

\subsection{Laser Doppler Velocimetry Principle}

In the differential mode, two coherent laser beams are crossed and focused to generate an ellipsoidal probe volume, in which the electromagnetic interferences lead to apparent dark and bright fringes \cite{Albrecht}.

\begin{figure}[h!]
\centering
\includegraphics[width=12cm]{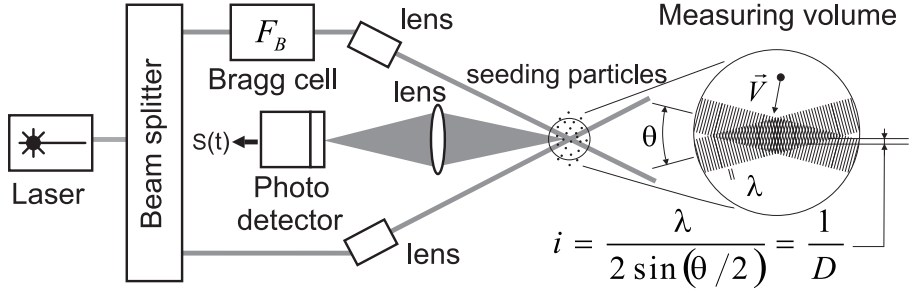}
\caption{\label{fig:setup} Optical setup of LDV system. When the particle $q$ crosses the measurement volume, the light is scattered in all directions and the burst signal $s_q(t)$ is collected by the photo detector. Data processing of  $s_q(t)$ allows then to estimate the mean flow and particle acoustic velocity.}
\end{figure}

The velocity $v_q(t)$ of the seeding particle denoted $q$ is related to the scattered optical field due to the Doppler effect. The light intensity scattered by the particle crossing the probe volume is modulated in amplitude and frequency. The frequency of modulation $F_q(t)$ is called Doppler frequency and is given by
\begin{equation}
\label{eq1}
F_q(t)=\frac{v_q(t)}{i}=\frac{2 v_q(t)}{\lambda_L} \sin (\theta /2),
\end{equation}
where $v_q(t)$ is the velocity of the particle along the $x$-axis, $i$ the fringe-spacing expressed as a function of the angle $\theta$ between the incoming laser beams and their optical wavelength $\lambda_L$ (Fig. \ref{fig:setup}).

The diffused light is collected by a receiving optics and is converted into an electrical signal by a photomultiplier (PM). This signal can then be modelled as \cite{Gazengel05}

\begin{equation}
\label{eq3}
s_q(t)=A_q(t) (M+\cos \phi_q (t)),
\end{equation}
where $M$ takes into account the positive sign of Cram\'er-Rao Bound (CRB) the light intensity. In (\ref{eq3}), the amplitude modulation linked to the normally distributed light intensity across the beam section is written as 
\begin{equation}
\label{eq:modulation}
A_q(t)= K_q e^{-(\beta d_q(t))^2},
\end{equation}
where $K_q$ is related to the laser beam, the PM sensitivity, the electronic amplification, the observation direction and the scattering efficiency of tracer $q$. Furthermore, $\beta$ is related to the probe geometry and  $d_q(t)$ is the projection of the time-varying particle displacement along the $x$-axis in the probe volume. Similarly, the phase modulation in (\ref{eq3}) is described by
\begin{equation}
\label{eq:phase}
\phi_q (t)= 2\pi \frac{d_q(t)}{i} + \phi_0,
\end{equation}
where $\phi_0$ is the initial phase due to optical setup. Furthermore, we denote $x_q(t)$ the signal such that
\begin{equation}
x_q(t)=s_q(t) + w(t),
\end{equation}
where $w(t)$ is the additive noise \cite{Valiere00}.

In order to avoid any ambiguity on the sign of the velocity, a Bragg cell tuned to frequency $F_B = 40$ MHz is used to shift the frequency of one of the lasers. The signal $s_q(t)$ is consequently written as 
\begin{equation}
\label{eq:Doppler_signal}
s_q(t)=A_q(t) (M+\cos ( 2 \pi F_B t + 2\pi d_q(t) / i + \phi_0 )).
\end{equation}

The offset component $M$ is then canceled by an high-pass filtering and the signal $s_q(t)$ is down shifted to zero thanks to a quadrature demodulation (QD) technique \cite{LeDuff}. The actual signal, called burst signal, can finally be written as

\begin{equation}
\label{eq:Doppler_signal2}
s_q(t)=A_q(t) \cos (2\pi d_q(t) / i + \phi_0 ).
\end{equation}

\subsection{Doppler Signal Modeling in Acoustics}

Considering only pure sine acoustic waves and supposing that the mean flow velocity is constant inside the probe volume, the projection along the $x$-axis of the velocity of a particle $q$ subjected jointly to the sine acoustic wave and the mean flow field can be expressed as
\begin{equation}
v_q(t) = v_{c,q} + V_{ac} \cos (2 \pi F_{ac} t + \phi_{ac}),
\label{eq:v_acoustic}
\end{equation}
where $v_{c,q}$ is the mean flow velocity of particle $q$, $V_{ac}$ and $\phi_{ac}$  are the amplitude and phase of the acoustic particle velocity and $F_{ac}$ is the known frequency of the pure sine acoustic excitation. The amplitude modulation of the burst signal (\ref{eq:modulation}) associated to the  particle $q$ may be written as
\begin{equation}
A_q(t)= K_q \exp [\beta (v_{c,q}(t-t_q) + \frac{V_{ac}}{2 \pi F_{ac}} \sin( 2 \pi F_{ac} t + \phi_{ac}) ) ]^2.
\label{eq:Aq_acoustic}
\end{equation}

Similarly, the phase modulation (\ref{eq:phase}) of the burst signal associated to the  particle $q$ is
\begin{equation}
\phi_q (t) = \frac{2\pi}{i}   v_{c,q} (t-t_q) + \frac{V_{ac}}{2 \pi F_{ac}} \sin( 2 \pi F_{ac} t + \phi_{ac}).
\label{eq:Phiq_acoustic}
\end{equation}

We note that the flow velocity $v_{c,q}$ can change from a particle $q$ to another while the acoustic parameters $v_{ac}$ and $\phi_{ac}$ are independent of $q$. Thus, when the acoustic wave is disturbed by a mean flow, assuming that the particles $q$ cross the measurement volume at different random central times $t_q$, without time overlapping between bursts $q$ and $q+1$, the Doppler signal can be written as
\begin{equation}
s(t)=\sum_q s_q(t) = A_D (t) \cos [\phi_D (t)],
\label{eq:Doppler_final}
\end{equation}
where the amplitude and phase respectively express as
\begin{equation}
A_D(t) = 
\left\lbrace
\begin{array}{l}
A_q (t), \mbox{ } t \in [t_q - T_q/2,t_q + T_q/2 ] \\
0, \mbox{ otherwise},
\end{array}
\right.
\end{equation}
and
\begin{equation}
\phi_D (t)=
\left\lbrace
\begin{array}{l}
\phi_q (t), \mbox{ } t \in [t_q - T_q/2,t_q + T_q/2 ]\\
0, \mbox{ otherwise}.
\end{array}
\right.
\end{equation}

\begin{figure}[h!]
\centering
\includegraphics[width=16cm]{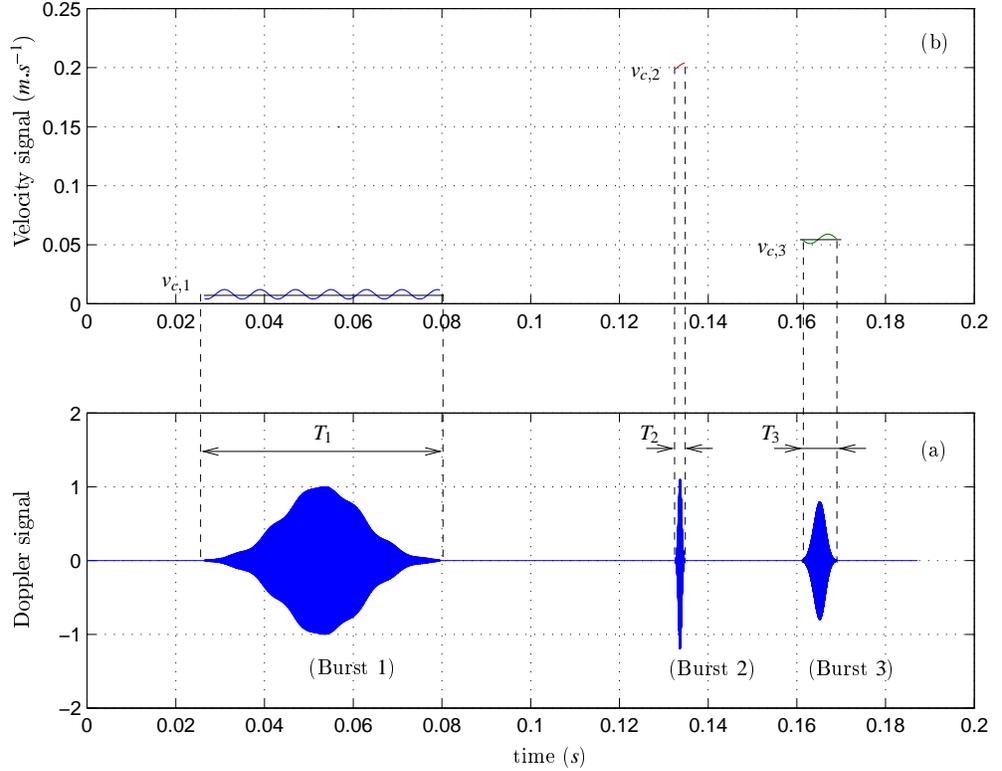}
\caption{\label{fig:ex_Doppler} (a) Example of a Doppler signal. (b) Associated velocity signal. Burst $1$ is associated with a low mean flow velocity corresponding to $N_{per}$ acoustic periods largely higher than $1$. Burst $2$ is associated with a high mean flow velocity corresponding to $N_{per}$ acoustic period largely lower than $1$. Burst $3$ is associated to a  mean flow velocity corresponding to $N_{per} \lesssim 1$ acoustic period.}
\end{figure}

Furthermore, the time of flight of the tracer $q$ is defined as \cite{Degroot}
\begin{equation}
\label{eq:tofq}
T_q=\frac{\sqrt{2} D_x}{v_{c,q}},
\end{equation}
where $D_x$ is the length of the probe volume in the $x$-axis and the associated number of acoustic periods is
\begin{equation}
\label{eq:Nper}
N_{per}=\frac{\sqrt{2} D_x}{v_{c,q}} F_{ac}.
\end{equation}
As expected, the fastest the particle crosses the probe volume, the lowest the time of flight and the number of acoustic periods. An example of a typical Doppler signal is shown on Fig. \ref{fig:ex_Doppler}(a), where the different particle times of flight are associated to different mean flow velocities.

\subsection{Doppler signal processing}

The aim of the signal processing developed after sampling the Doppler signal is to estimate jointly and burst-by-burst the acoustic particle velocity (amplitude $V_{ac}$ and phase $\phi_{ac}$) and the mean flow velocity $v_{c,q}$. This procedure is usually split into two stages. After a detection procedure \cite{Degroot}, a frequency demodulation of the Doppler signal $s(t)$ is performed by using a time-frequency transform to estimate the instantaneous frequency $F_q(t)$, or equivalently (\ref{eq1}) the velocity signal $v_q(t)$, burst by burst \cite{Valeau04}. Note that the detector selects only bursts corresponding to one tracer in the measurement volume. Secondly, the data processing of the estimated velocity signal $\hat{v}_q(t)$ allows to obtain both components of the acoustic and mean flow velocities for each burst. This first stage is described in this subsection and the second one (data LMS-based processing) is explained in the section \ref{LMSA}.

According to (\ref{eq1}), the velocity signal associated to the particle $q$ expresses as
\begin{equation}
\label{eq:vq}
v_q(t) = i F_q (t),
\end{equation} 
where $i$ is the fringe spacing. Thus, the problem consists to estimate the mean value $\hat{v}_{c,q}$, the amplitude $\hat{V}_{ac}$ and the phase $\hat{\phi}_{ac}$ of the estimated velocity signal associated to each burst $q$ from the actual noisy burst signal $x_q(t)$. Fig. \ref{fig:ex_Doppler} shows an example of an noiseless simulated Doppler signal (a) and the associated velocity signal for three non-overlapping bursts (b).

In section \ref{sec:CRB}, the Cram\'er Rao bound of the problem is calculated. Then, a method based on a least mean square algorithm is presented in section \ref{LMSA} and is applied to simulated velocity signals $v_q(t)$ in section \ref{sec:processing}. 

\section{CRB calculation}
\label{sec:CRB}

We recall that the Cram\'er-Rao bound (CRB) gives the lowest bound of the variance an unbiased estimator may reach (if it exists) \cite{Kay}. As explained in \cite{Kay}, the CRB alerts us to the physical impossibility of finding an unbiased estimator whose variance is less than the bound.
In the case of single tone signals, CRB were calculated by Rife and Boorstyn \cite{Rife} in 1974. The CRB of LDV signals were also studied in the case of fluid mechanics \cite{Besson}-\cite{Shu}. In the case of sine acoustic excitation, the CRB of LDV signal were also studied by Le Duff \cite{LeDuff}.

We focus here on the problem of calculating the Cram\'er-Rao bound (CRB) of the following problem. The velocity data are assumed to be such that 
\begin{equation}
\label{eqdp5}
u[n]=v[n;{\mathbf{\theta}}]+w[n],
\end{equation}
for $n \in [n_0,n_1]$, where $w[n]$ is the WGN, $w[n]\sim {\cal N}(0,\sigma^2)$, the data being modeled according to 
\begin{equation}
\label{eqdp1}
v[n;{\mathbf{\theta}}]=v_c+V_{ac} \cos(2\pi f_{ac} n+ \phi_{ac}),
\end{equation}
where $f_{ac} = F_{ac}/F_s$, $F_s$ being the sampling frequency, $v_c \equiv v_{c,q}$, and where the unknown parameters are gathered in 
\begin{equation}
\label{eqdp2}
{\mathbf{\theta}}={[v_c \;\;\; V_{ac} \;\;\; \phi_{ac}]}^T.
\end{equation}
We furthermore suppose that $f_{ac} \neq 0$ and $f_{ac} \neq \frac{1}{2}$. 

\subsection{Cram\'er-Rao bound (CRB) for one burst}
\label{CRB1}

The CRB is given by the inverse of the Fisher information matrix ${\mathbf{J}}({\mathbf{\theta}})$, $\mbox{CRB}({\mathbf{\theta}})={{\mathbf{J}}({\mathbf{\theta}})}^{-1}$, where the Fisher information matrix is given by \cite{Kay}

\begin{equation}
\label{eqdp6}
{\mathbf{J}}({\mathbf{\theta}})_{kl}=\frac{1}{\sigma^2}
\sum_{n=n_0}^{n_1}
\frac{\partial v[n;\theta]}{\partial \theta_k}
\frac{\partial v[n;\theta]}{\partial \theta_l},
\end{equation}
for $k,l \in [1,3]$, for ${\mathbf{\theta}}={[v_c \;\;\; V_{ac} \;\;\; \phi_{ac}]}^T$. The derivatives in (\ref{eqdp6}), according to (\ref{eqdp1}), lead to
\begin{equation}
\label{eqdp7}
{\mathbf{J}}({\mathbf{\theta}})=
\frac{1}{\sigma^2}
\left(
\begin{array}{ccc}
N & \frac{\mbox{cos}(\beta) \mbox{sin}(\gamma N)}{\mbox{sin}(\gamma)} & -\frac{V_{ac}\mbox{sin}(\beta) \mbox{sin}(\gamma N)}{\mbox{sin}(\gamma)} \\
\frac{\mbox{cos}(\beta) \mbox{sin}(\gamma N)}{\mbox{sin}(\gamma)} & \frac{N}{2}+\frac{\mbox{cos}(2\beta)\mbox{sin}(2\gamma N)}{2\mbox{sin}(2\gamma)} & -\frac{V_{ac}\mbox{sin}(2\beta)\mbox{sin}(2\gamma N)}{2\mbox{sin}(2\gamma)} \\
- \frac{V_{ac}\mbox{sin}(\beta)\mbox{sin}(\gamma N)}{\mbox{sin}(\gamma)} & -\frac{V_{ac}\mbox{sin}(2\beta)\mbox{sin}(2\gamma N)}{2\mbox{sin}(2\gamma)} & \frac{N V_{ac}^2}{2}-\frac{V_{ac}^2\mbox{cos}(2\beta)\mbox{sin}(2\gamma N)}{2\mbox{sin}(2\gamma)} 
\end{array}
\right),
\end{equation}
 where $N=n_1-n_0+1$, and where 
\begin{equation}
\label{eqdp7bis}
\gamma = \pi f_{ac},
\end{equation}
\begin{equation}
\label{eqdp7ter}
\beta = 2\pi f_{ac} n_0+\pi f_{ac} (N-1) +\phi_{ac}.
\end{equation}

We define the linear signal-to-noise ratio (SNR) of the velocity signal as
\begin{equation}
\label{eq:sigma}
\mbox{SNR}= \frac{V_{ac}^2}{2 \sigma^2},
\end{equation}
and we then have upon inversion
\begin{equation}
\label{eqdp8}
\mbox{var}(v_c) \geq \mbox{CRB}(v_c)=
\frac{V_{ac}^2}{4 \mbox{ SNR}}
\frac{N^2-{\bigg(\frac{\mbox{sin}(2\gamma N)}{\mbox{sin}(2 \gamma)}\bigg)}^2}
{\frac{N^3}{2}-\frac{N}{2}{\bigg(\frac{\mbox{sin}(2\gamma N)}{\mbox{sin}(2 \gamma)}\bigg)}^2-N {\bigg(\frac{\mbox{sin}(\gamma N)}{\mbox{sin}(\gamma)}\bigg)}^2+
\frac{\mbox{sin}(2\gamma N)}{\mbox{sin}(2 \gamma)}{\bigg(\frac{\mbox{sin}(\gamma N)}{\mbox{sin}(\gamma)}\bigg)}^2},
\end{equation}
\begin{equation}
\label{eqdp9}
\mbox{var}(V_{ac}) \geq \mbox{CRB}(V_{ac})=
\frac{V_{ac}^2}{2 \mbox{ SNR}}
\frac{N^2-N \mbox{cos}(2\beta)\frac{\mbox{sin}(2\gamma N)}{\mbox{sin}(2 \gamma)}-
2{\mbox{sin}}^2(\beta){\bigg(\frac{\mbox{sin}(\gamma N)}{\mbox{sin}(\gamma)}\bigg)}^2}
{\frac{N^3}{2}-\frac{N}{2}{\bigg(\frac{\mbox{sin}(2\gamma N)}{\mbox{sin}(2 \gamma)}\bigg)}^2-N {\bigg(\frac{\mbox{sin}(\gamma N)}{\mbox{sin}(\gamma)}\bigg)}^2+
\frac{\mbox{sin}(2\gamma N)}{\mbox{sin}(2 \gamma)}{\bigg(\frac{\mbox{sin}(\gamma N)}{\mbox{sin}(\gamma)}\bigg)}^2},
\end{equation}
\begin{equation}
\label{eqdp10}
\mbox{var}(\phi_{ac}) \geq \mbox{CRB}(\phi_{ac})=
\frac{1}{2 \mbox{ SNR}}
\frac{N^2+N \mbox{cos}(2\beta)\frac{\mbox{sin}(2\gamma N)}{\mbox{sin}(2 \gamma)}-
2{\mbox{cos}}^2(\beta){\bigg(\frac{\mbox{sin}(\gamma N)}{\mbox{sin}(\gamma)}\bigg)}^2}
{\frac{N^3}{2}-\frac{N}{2}{\bigg(\frac{\mbox{sin}(2\gamma N)}{\mbox{sin}(2 \gamma)}\bigg)}^2-N {\bigg(\frac{\mbox{sin}(\gamma N)}{\mbox{sin}(\gamma)}\bigg)}^2+
\frac{\mbox{sin}(2\gamma N)}{\mbox{sin}(2 \gamma)}{\bigg(\frac{\mbox{sin}(\gamma N)}{\mbox{sin}(\gamma)}\bigg)}^2}.
\end{equation}

\subsection{Cram\'er-Rao Bound (CRB) for $N_b$ bursts}

We now assume that the algorithm developed in \ref{CRB1} is used for estimating the unknown parameters ${\mathbf{\theta}}={[v_c \;\;\; V_{ac} \;\;\; \phi_{ac}]}^T$, in the case of $N_b$ bursts. The main difference between this problem and the one developed above is that the index $n_0$ is not anymore a constant, but might be modeled as a discrete random variable, uniformly distributed in $[0,N_{ac}]$, where $N_{ac}=\mbox{nint}\bigg[F_s / F_{ac} \bigg]$, nint[] being the nearest integer.
As a consequence, the discrete random variable $\beta$ given in (\ref{eqdp7ter}), which appears in (\ref{eqdp9}-\ref{eqdp10}) is uniformly distributed in $[\pi (N-1) f_{ac}+ \phi_{ac}, \pi (N-1) f_{ac}+\phi_{ac}+2\pi]$. Averaging the terms linked to $\beta$ in  (\ref{eqdp9}) and (\ref{eqdp10}) consequently leads to 
\begin{equation}
<\mbox{cos}(2\beta)>=<\mbox{sin}(2\beta)>=0,
\end{equation}
and
\begin{equation}
<\mbox{cos}^2(\beta)>=<\mbox{sin}^2(\beta)>=\frac{1}{2}.
\end{equation}

This finally yields
\begin{equation}
\label{eqdp11}
\mbox{var}(V_{ac}) \geq \mbox{CRB}(V_{ac})=\frac{V_{ac}^2}{2 \mbox{ SNR}}
\frac{N^2-{\bigg(\frac{\mbox{sin}(\gamma N)}{\mbox{sin}(\gamma)}\bigg)}^2}
{\frac{N^3}{2}-\frac{N}{2}{\bigg(\frac{\mbox{sin}(2\gamma N)}{\mbox{sin}(2 \gamma)}\bigg)}^2-N {\bigg(\frac{\mbox{sin}(\gamma N)}{\mbox{sin}(\gamma)}\bigg)}^2+
\frac{\mbox{sin}(2\gamma N)}{\mbox{sin}(2 \gamma)}{\bigg(\frac{\mbox{sin}(\gamma N)}{\mbox{sin}(\gamma)}\bigg)}^2}
\end{equation}
and
\begin{equation}
\label{eqdp12}
\mbox{var}(\phi_{ac}) \geq \mbox{CRB}(\phi_{ac})=\frac{1}{2 \mbox{ SNR}}
\frac{N^2-{\bigg(\frac{\mbox{sin}(\gamma N)}{\mbox{sin}(\gamma)}\bigg)}^2}
{\frac{N^3}{2}-\frac{N}{2}{\bigg(\frac{\mbox{sin}(2\gamma N)}{\mbox{sin}(2 \gamma)}\bigg)}^2-N {\bigg(\frac{\mbox{sin}(\gamma N)}{\mbox{sin}(\gamma)}\bigg)}^2+
\frac{\mbox{sin}(2\gamma N)}{\mbox{sin}(2 \gamma)}{\bigg(\frac{\mbox{sin}(\gamma N)}{\mbox{sin}(\gamma)}\bigg)}^2}.
\end{equation}

In the following, we use the expressions (\ref{eqdp8}) for $v_c$ and (\ref{eqdp11}) for $V_{ac}$ for studying the CRB of the problem. We recall that $N$ depends on $v_c$ (\ref{eq:nq}). As a consequence, the CRB of $v_c$ (\ref{eqdp8}) and the CRB of $V_{ac}$ (\ref{eqdp11}) both depend on $v_c$ and $V_{ac}$, while the CRB of $\phi_{ac}$ (\ref{eqdp12}) is independent of $V_{ac}$.

\subsection{Asymptotic behavior of Cram\'er-Rao Bound (CRB)}
\label{CRB3}

In Appendix \ref{appendix2}, we give the expressions of the asymptotic CRB of $\mathbf{\theta}$, for both cases $2\gamma N \ll 1$ ($N_{per} \ll 1/(2 \pi)$) and $2\gamma N \gg 1$ ($N_{per} \gg 1/(2 \pi)$). 

In the asymptotic case  $2\gamma N \ll 1$, we prove (\ref{eqapp2_3ter}-\ref{eqapp2_4ter}) that the relative variance of $v_c$ and $V_{ac}$ are 
\begin{equation}
\label{eqapp2_3bis}
\frac{\mbox{var}(v_c)}{v_c^2} \geq \frac{\mbox{CRB}(v_c)}{v_c^2} = \frac{1}{\mbox{SNR}} \frac{45}{\pi^4 2^{7/2}}\frac{1}{D_x^5 F_s}\frac{v_c^3 V_{ac}^2}{F_{ac}^4},
\end{equation}
\begin{equation}
\label{eqapp2_4bis}
\frac{\mbox{var}(V_{ac})}{V_{ac}^2} \geq \frac{\mbox{CRB}(V_{ac})}{V_{ac}^2} = \frac{1}{\mbox{SNR}}  \frac{45}{\pi^4 2^{9/2}}\frac{1}{D_x^5 F_s}\frac{v_c^5}{F_{ac}^4}.
\end{equation}

Both CRBs of $v_c$ and $V_{ac}$ are proportional to $v_c^5 V_{ac}^2$ and inversely proportional to $F_{ac}^4$. Consequently, doubling the mean flow velocity yields an $15$ dB increase of the variance of both $v_c$ and $V_{ac}$. Similarly, doubling the amplitude of the acoustic particle velocity $V_{ac}$ leads to a $6$ dB increase of the variance of both $v_c$ and $V_{ac}$. Lastly, doubling the frequency of the pure sine acoustic wave leads to a $12$ dB decrease of the variance of both $v_c$ and $V_{ac}$. We also note that doubling the length of the probe volume $D_x$ yields a $15$ dB decrease of the variance of both $v_c$ and $V_{ac}$.

In the asymptotic case $2 \gamma N \gg 1$, we prove that (\ref{eqapp3_6bis}-\ref{eqapp3_7bis})
\begin{equation}
\label{eq:var_vc_asymp}
\mbox{var}(v_c) \geq \frac{1}{\mbox{SNR}} \frac{1}{2^{3/2} D_x F_s} v_c V_{ac}^2,
\end{equation}
and
\begin{equation}
\label{eq:var_vac_asymp}
\mbox{var}(V_{ac}) \geq \frac{1}{\mbox{SNR}} \frac{1}{\sqrt{2} D_x F_s} v_c V_{ac}^2.
\end{equation}

Thanks to the exact (\ref{eqdp8}, \ref{eqdp11}, \ref{eqdp12}) and asymptotic (\ref{eqapp2_3bis}-\ref{eq:var_vac_asymp}) expressions of the CRB, the minimum uncertainties linked to the velocity estimations (acoustic and mean flow velocities) are completely known. In section (\ref{LMSA}), the LMS-based algorithm is introduced. It is then applied in section (\ref{sec:processing}) to simulated data in order to be compared with the CRB.

\section{Least mean square algorithm}
\label{LMSA}

From a practical point of view, the actual velocity signal is uniformly sampled. Consequently, the number of samples $N_q$ associated to the particle $q$ is derived from (\ref{eq:tofq}), as 
\begin{equation}
\label{eq:nq}
N_q=\frac{\sqrt{2} D_x F_s}{v_{c,q}},
\end{equation}
and the associated number of acoustic periods (\ref{eq:Nper}) is now defined as
\begin{equation}
\label{eq:np}
N_{per}=\frac{\sqrt{2} D_x}{v_{c,q}} F_{ac}.
\end{equation}

The sine-wave fit is then solved by minimizing the cost function $V({\mathbf{\theta}})$,
\begin{equation}
\label{eqdp3}
V({\mathbf{\theta}})=\frac{1}{N} \sum_{n=n_0}^{n_1}
{(u[n]-v[n;{\mathbf{\theta}}])}^2,
\end{equation}
with respect to the unknown parameters $\mathbf{\theta}$ (\ref{eqdp2}), where $u[n]$ and $v[n;{\mathbf{\theta}}]$ are respectively given by (\ref{eqdp5}) and (\ref{eqdp1}), and where $N=n_1-n_0+1$. In the Appendix 1, the equations (\ref{eqapp1_5}-\ref{eqapp1_7}) respectively give the expression of $v_c$, $a_{ac}=V_{ac} \cos (\phi_{ac})$ and $b_{ac}=V_{ac} \sin (\phi_{ac})$ as a function of $\mathbf{u}$ and $f_{ac}$. Once $a_{ac}$ and  $b_{ac}$ are estimated, the unknown acoustical parameters of  $\mathbf{\theta}$ express as 
\begin{equation}
\label{eqdp4}
\left\lbrace
\begin{array}{l}
\hat{V}_{ac}=\sqrt{\hat{a}_{ac}^2+\hat{b}_{ac}^2}, \\
\hat{\phi}_{ac}=\mbox{atan} \frac{\hat{b}_{ac}}{\hat{a}_{ac}}.
\end{array}
\right.
\end{equation}

\section{Numerical results and discussion}
\label{sec:processing}

In this section, we compare the CRB with the LMS-based algorithm developed in \ref{LMSA}. According to the values of the acoustic and mean flow velocities to be analyzed, the following values for $F_{ac}$ and $V_{ac}$ are chosen : 
\begin{eqnarray}
\label{eqnr1}
F_{ac} & \in & [125 \;\;\; 250 \;\;\; 500 \;\;\; 1000 \;\;\; 2000 \;\;\; 4000] \mbox{ Hz}, \\
V_{ac} & \in & [0.05 \;\;\; 1.58 \;\;\; 50] \mbox{ mms}^{-1}.
\end{eqnarray}

The phase $\phi_{ac}$ is supposed to be equal to $\pi/4$, and we use an adimensional parameter $\alpha_v$ for the value of $v_c$, such that
\begin{equation}
\label{eqnr2}
\alpha_v=\frac{V_{ac}}{v_c} \in [0.01 \;\;\; 0.05 \;\;\; 0.1 \;\;\; 0.5 \;\;\; 1].
\end{equation}

For each numerical simulation, the sampling frequency is $F_s=350$ kHz, the probe volume length along the $x-$axis is $D_x=0.1$ mm and $10000$ bursts are analyzed. The simulator is performed by Matlab.

\begin{figure}[h!]
\centering
\includegraphics[width=15cm]{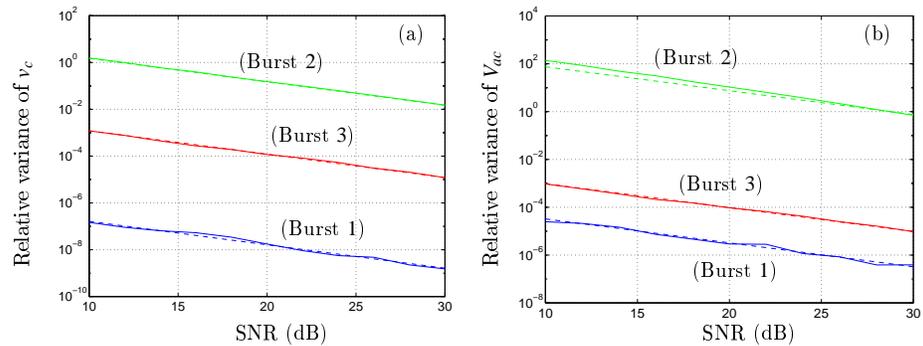}
\caption{\label{figcrb125} Comparison of the relative variances of $v_c$ (a) and $V_{ac}$ (b) estimated by a LMS algorithm (continuous) with the theoretical CRB (dashed), for $F_{ac}=125$ Hz. Bursts ($1-3$) refer to Fig. \ref{fig:ex_Doppler}. (Burst $1$)~:~$V_{ac}=1.58$ mm.s$^{-1}$, $\alpha_v=0.1$, $v_{c}=15.8$ mm.s$^{-1}$.
(Burst $2$)~:~$V_{ac}=50$ mm.s$^{-1}$, $ \alpha_v=0.1$, $v_{c}=500$ mm.s$^{-1}$.
(Burst $3$)~:~$V_{ac}=50$ mm.s$^{-1}$, $ \alpha_v=1$, $v_{c}=50$ mm.s$^{-1}$.}
\end{figure}

\begin{figure}[h!]
\centering
\includegraphics[width=15cm]{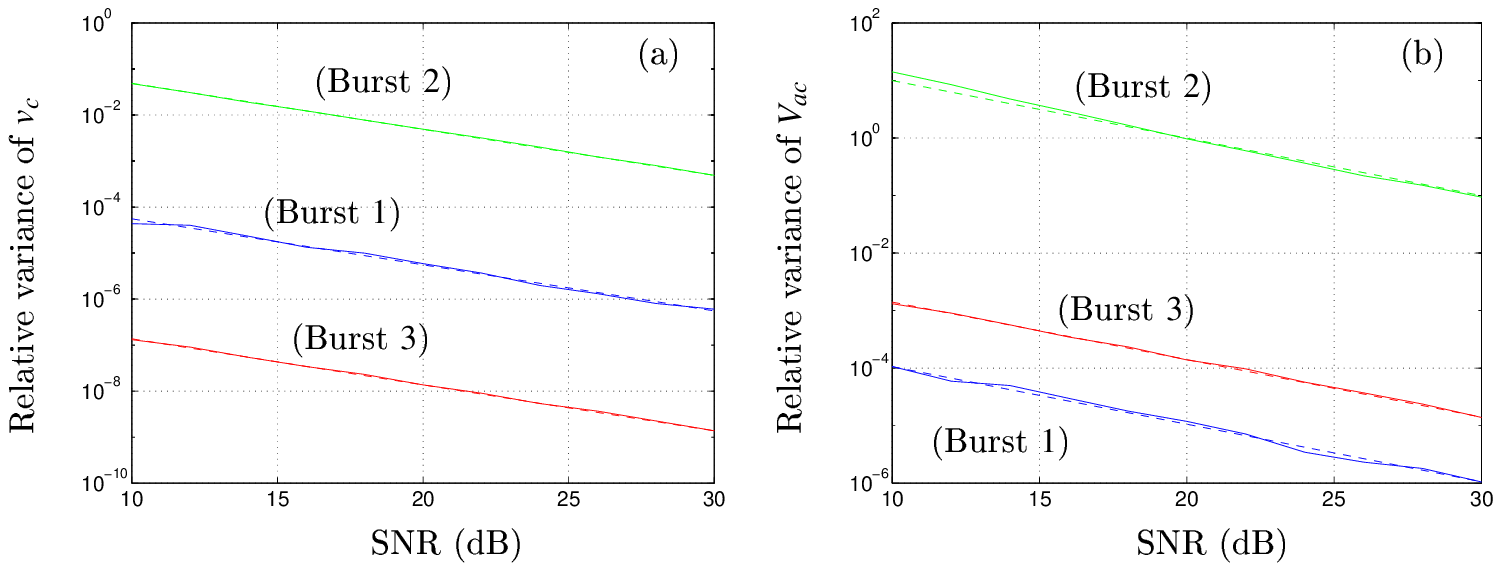}
\caption{\label{figcrb500} Comparison of the relative variances of $v_c$ (a) and $V_{ac}$ (b) estimated by a LMS algorithm (continuous) with the theoretical CRB (dashed), for $F_{ac}=500$ Hz. Bursts ($1-3$) refer to Fig. \ref{fig:ex_Doppler}. (Burst $1$)~:~$V_{ac}=50$ mm.s$^{-1}$, $ \alpha_v=1$, $v_{c}=50$ mm.s$^{-1}$.
(Burst $2$)~:~$V_{ac}=50$ mm.s$^{-1}$, $ \alpha_v=0.05$, $v_{c}=1000$ mm.s$^{-1}$.
(Burst $3$)~:~$V_{ac}=1.58$ mm.s$^{-1}$, $ \alpha_v=0.1$, $v_{c}=15.8$ mm.s$^{-1}$.}
\end{figure}

\begin{figure}[h!]
\centering
\includegraphics[width=15cm]{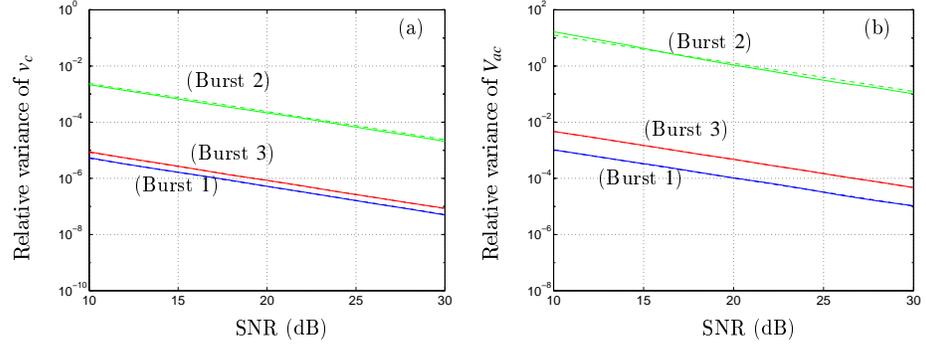}
\caption{\label{figcrb4000} Comparison of the relative variances of $v_c$ (a) and $V_{ac}$ (b) estimated by a LMS algorithm (continuous) with the theoretical CRB (dashed), for $F_{ac}=4000$ Hz. Bursts ($1-3$) refer to Fig. \ref{fig:ex_Doppler}. (Burst $1$)~:~$V_{ac}=50$ mm.s$^{-1}$, $ \alpha_v=0.1$, $v_{c}=500$ mm.s$^{-1}$.
(Burst $2$)~:~$V_{ac}=50$ mm.s$^{-1}$ , $\alpha_v=0.01$, $v_{c}=5000$  mm.s$^{-1}$. 
(Burst $3$)~:~$V_{ac}=50$ mm.s$^{-1}$, $ \alpha=0.05$, $v_{c}=1000$ mm.s$^{-1}$. }
\end{figure}

Fig. \ref{figcrb125} to Fig. \ref{figcrb4000} show typical results of the relative variances $\mbox{var}(v_c)/v_c^2$ (a) and $\mbox{var}(V_{ac})/V_{ac}^2$ (b), for the different values of $F_{ac}$ with comparison to the theoretical CRB of $v_c$ (\ref{eqdp8}) and of $V_{ac}$ (\ref{eqdp9}). Each figure is related to a given value of $F_{ac}$. Furthermore, for each value of $F_{ac}$, three sets of signals are analyzed, each set corresponding to one of the bursts of Fig. \ref{fig:ex_Doppler}, respectively $N_{per} \gg 1$ $(\mbox{Burst } 1)$, $N_{per} > 1$ $(\mbox{Burst } 2)$ and $N_{per} \lesssim 1$ $(\mbox{Burst } 3)$, where $N_{per}$ is the number of acoustic periods, $N_{per}=N f_{ac}$.

First of all, the LMS-based estimator is near the theoretical CRB, so that we can maintain that this estimator is efficient. Moreover, the relative variance of $v_c$ is weaker than the one of $V_{ac}$ (except for $\alpha_v=1$). Indeed, for values of $v_c$ and $V_{ac}$ such that $\alpha_v=V_{ac}/v_c \leq 1/\sqrt{2}$ as $\mbox{CRB}(v_c) \simeq \mbox{CRB}(V_{ac})/2$, we have

\begin{equation}
\label{eq:CRB_vc}
\frac{\mbox{CRB}(v_c)}{v_c^2} \leq \frac{\mbox{CRB}(V_{ac})}{V_{ac}^2}.
\end{equation}

Moreover, the values of the relative variances of $v_c$ and $V_{ac}$ drastically depend on the values of $v_c$, $V_{ac}$ and $F_{ac}$ as shown by (\ref{eqdp8}) and (\ref{eqdp11}).
\begin{itemize}
\item[-] For $N_{per} > 1$ (Burst $1$), the relative variances of  $v_c$ and $V_{ac}$ are respectively in $[-70,-42]$ dB and $[-52,-30]$ dB. 
 
Such estimations may consequently be considered as very accurate.
\item[-] For $N_{per} \ll 1$ (Burst $2$), the relative variances of $v_c$ and $V_{ac}$ are respectively in $[-27,1]$ dB and $[10,20]$ dB. The estimation of $v_c$ is accurate enough for low values of $v_c^3V_{ac}^2/F_{ac}^4$ (\ref{eqapp2_3bis}), while the estimation of $V_{ac}$ is clearly unacceptable, whatever the parameters $v_c$, $V_{ac}$ and $F_{ac}$ (\ref{eqapp2_4bis}).
\item[-] For $N_{per} \lesssim 1$ (Burst $3$), the relative variances of $v_c$ and $V_{ac}$ are respectively in $[-70,-30]$ dB and $[-30,-23]$ dB. Such estimations may also be considered as very accurate.
\end{itemize}

Furthermore, for velocity signals with time length largely lower than one acoustic period, we can use the asymptotic case expression of CRB of $v_c$ and $V_{ac}$. Giving a maximum value of relative error, respectively $E_{v_c}$ for $v_c$ and $E_{v_{ac}}$ for $V_{ac}$, we consequently have

\begin{equation}
\frac{\mbox{CRB}(v_c)}{v_c^2} = \frac{1}{\mbox{SNR}} \frac{45}{\pi^4 2^{7/2}} \frac{1}{D_x^5 F_s} \frac{v_c^3 V_{ac}^2}{F_{ac}^4} \leq E_{v_c}
\end{equation}
and
\begin{equation}
\frac{\mbox{CRB}(V_{ac})}{V_{ac}^2} = \frac{1}{\mbox{SNR}} \frac{45}{\pi^4 2^{9/2}} \frac{1}{D_x F_s} \frac{v_c^5}{F_{ac}^4} \leq E_{v_{ac}}.
\end{equation}

As a consequence, for a given set of setup known parameters $D_x$, $F_s$ and $F_{ac}$, we may give the maximum values $v_c^3 V_{ac}^2$ and $v_c^5$ have to reach for yielding an error less than respectively $E_{v_c}$ and $E_{v_{ac}}$.\\

Lastly, from the expressions of the CRB of $v_c$ (\ref{eqdp8}) and $V_{ac}$ (\ref{eqdp11}), we can calculate the number of acoustic periods the time length of the velocity signals may have, for leading to an error less than a given value $E$. Tables \ref{tab:125_pourcent}-\ref{tab:4000_pourcent} give a summary of such results. Each table corresponds to a given burst of Fig. \ref{fig:ex_Doppler}.

\begin{table}[h!]
\caption{\label{tab:125_pourcent} Number of acoustic periods $N_{per}$ (for $v_c$ and for $V_{ac}$) leading to an error less than $E(\%)$ for $F_{ac}=125$ Hz and $V_{ac}= 50$ mm.s$^{-1}$.}
\centering
\begin{tabular}[t]{||l|l|l|l|l|l|l|l|l|l||}
\hline
SNR (dB) & \multicolumn{3}{c|}{$10$} & \multicolumn{3}{c|}{$20$} & \multicolumn{3}{c||}{$30$} \\
\hline
$E$ (\%) & $0.1$ & $1$ & $10$  & $0.1$ & $1$ & $10$ & $0.1$ & $1$ & $10$ \\
\hline \hline
$N_{per}\;(v_c)$ & $ \geq 0.8$ & $\geq 0.3$ & $\geq 0.2$ & $\geq 0.45$ & $\geq 0.2$ & $\geq 0.09$ & $\geq 0.3$ & $ \geq 0.12$ & $\geq 0.05$ \\
$N_{per}\;(V_{ac})$ & $\gg 10$ & $\geq 0.75$ & $\geq 0.25$ & $\gg 5 $ & $\geq 0.5$ & $\geq 0.2$ & $\geq 0.8$ & $\geq 0.25$ & $\geq 0.1$ \\
\hline
\end{tabular}
\end{table}

\begin{table}[h!]
\caption{\label{tab:500_pourcent} Number of acoustic periods $N_{per}$ (for $v_c$ and for $V_{ac}$) leading to an error less than $E(\%)$ for $F_{ac}=500$ Hz and $V_{ac}= 1.58$ mm.s$^{-1}$.}
\centering
\begin{tabular}[t]{||l|l|l|l|l|l|l|l|l|l||}
\hline
SNR (dB) & \multicolumn{3}{c|}{$10$} & \multicolumn{3}{c|}{$20$} & \multicolumn{3}{c||}{$30$} \\
\hline
$E$ (\%) & $0.1$ & $1$ & $10$  & $0.1$ & $1$ & $10$ & $0.1$ & $1$ & $10$ \\
\hline \hline
$N_{per}\;(v_c)$ & $\geq 0.4$ & $\geq 0.16$ & $\geq 0.06$ & $\geq 0.25$ & $ \geq 0.1$ & $ \geq 0.04$ & $\geq 0.18$ & $\geq 0.06$ & $ \geq 0.02$ \\
$N_{per}\;(V_{ac})$ & $\gg 10$ & $\geq 6$ & $\geq 0.4$ & $ \gg 10$ & $\geq 0.75$ & $ \geq 0.25$ & $ \gg 5$ & $ \geq 0.9$ & $\geq 0.15$ \\
\hline
\end{tabular}
\end{table}

\begin{table}[h!]
\caption{\label{tab:4000_pourcent} Number of acoustic periods $N_{per}$ (for $v_c$ and for $V_{ac}$) leading to an error less than $E(\%)$ for $F_{ac}=4000$ Hz and $V_{ac}=50$ mm.s$^{-1}$.}
\centering
\begin{tabular}[t]{||l|l|l|l|l|l|l|l|l|l||}
\hline
SNR (dB) & \multicolumn{3}{c|}{$10$} & \multicolumn{3}{c|}{$20$} & \multicolumn{3}{c||}{$30$} \\
\hline
$E$ (\%) & $0.1$ & $1$ & $10$  & $0.1$ & $1$ & $10$ & $0.1$ & $1$ & $10$ \\
\hline \hline
$N_{per}\;(v_c)$ & $\geq 0.6$ & $\geq 0.25$ & $ \geq 0.1$ & $ \geq 0.4$ & $ \geq 0.15$ & $ \geq 0.05$ & $ \geq 0.25$ & $\geq 0.1$ & $ \geq 0.03$ \\
$N_{per}\;(V_{ac})$ & $\gg 20$ & $\gg 5$ & $ \geq 0.6$ & $\gg 10$ & $ \geq 4.5$ & $ \geq 0.04$ & $\gg 10$ & $ \geq 0.7$ & $ \geq 0.25$ \\
\hline
\end{tabular}
\end{table}

For example, table \ref{tab:125_pourcent} may be read as follows. To obtain a relative error for $v_c$ less than $0.1$ \% for SNR$=10$ dB, the minimum number of acoustic period for the velocity signal is $0.8$. In the same way, to obtain a relative error for $V_{ac}$ less than $1$ \%  for SNR$=20$ dB, the minimum number of acoustic period for the velocity signal is $0.5$. Tables \ref{tab:500_pourcent}-\ref{tab:4000_pourcent} give the minimum number of acoustic periods the velocity signal should have for obtaining relative errors less than $0.1$ \%, $1$ \% and $10$ \%, for SNR equals to $10$ dB, $20$ dB and $30$ dB for $500$ Hz and $4000$ Hz respectively.

As expected, the mean flow velocity $v_c$ is estimated with a great accuracy from a very low number of acoustic period. For example, to obtain a relative error of $1$ \% for $v_c$, the number of acoustic period is always less than $0.3$ whatever the SNR, $F_{ac}$ and $V_{ac}$. On contrary, the results for the estimation of the acoustic velocity are much more contrasted. For a SNR of $30$ dB, the estimation of $V_{ac}$ associated with a relative error less than $1$ \% is possible for a number of acoustic period $N_{per}  > 0.9$. But, when the SNR is less than $30$ dB, the number of acoustic periods associated with a relative error less than $1$ \% may be largely bigger than $1$. 

The tables also show the influence of the acoustic frequency on the estimation of the particle acoustic velocity. The higher the frequency, the higher the number of acoustic periods for an accurate estimation of $V_{ac}$. For a relative error equals to $10$ \%, the number of acoustic periods are the same whatever the frequency. But for a relative error equals to $1$ \% or $0.1$ \% the estimation of the particle acoustic velocity is easier for a low frequency. On contrary, the influence of the frequency on the estimation of the mean flow velocity is the opposite. The higher the frequency, the lower the number of the acoustic periods for an accurate estimation.

With regard to the results of these study, a three-steps new approach can be proposed to improve the estimation of the acoustic particle velocity in presence of mean flow. The first step consists in the estimation of the mean flow velocity for each burst with the least mean square (LMS) algorithm. Then, the estimation of the mean flow velocity may be subtracted from the velocity signal. Finally, a "rotating machinery" technique associated with a synchronous detection allows to estimate the acoustic particle velocity with a great accuracy \cite{Gazengel03}.

\section{Conclusion}

A new method for estimating jointly the acoustic particle and the mean flow velocities from a LDV signal is presented. It is based on the least mean square (LMS) algorithm and it performs well in the estimation of the velocities.  The performance of the method has been investigated by means of numerical tests and the results of the simulation have been compared to the Cram\'er-Rao bounds of the associated problem. It is shown that the LMS-based estimator is near the theoretical CRB, so that the estimator is efficient.

\newpage
\appendices
\section{Derivation of LMS problem}
\label{appendix1}

Inserting (\ref{eqdp1}) into (\ref{eqdp3}) leads to

\begin{equation}
\label{eqapp1_1}
V(\theta)=\frac{1}{N} \sum_{n=n_0}^{n_1}
{\bigg(u_n-(v_c+a_{ac} \mbox{cos}(2\pi f_{ac} n)+b_{ac} \mbox{sin}(2\pi f_{ac} n)\bigg)}^2,
\end{equation}
where 
\begin{equation}
\label{eqapp1_2}
\left\lbrace
\begin{array}{l}
a_{ac}=V_{ac} \mbox{cos}(\phi_{ac}), \\
b_{ac}=V_{ac} \mbox{sin}(\phi_{ac}).
\end{array}
\right.
\end{equation}

Solving the following linear problem

\begin{equation}
\label{eqapp1_3}
\left\lbrace
\begin{array}{l}
\frac{\partial V(\mathbf{\theta})}{\partial v_c} = 0, \\
\frac{\partial V(\mathbf{\theta})}{\partial V_{ac}} = 0, \\
\frac{\partial V(\mathbf{\theta})}{\partial \phi_{ac}} = 0,
\end{array}
\right.
\end{equation}
allows to write analytically the unknown parameters. In the following, we note

\begin{eqnarray}
\label{eqapp1_4}
D & = & \frac{1}{N^2} \sum_{n=n_0}^{n=n_1}\mbox{cos}^2 (2{\pi}f_{ac} n)\sum_{n=n_0}^{n=n_1}\mbox{sin}^2(2{\pi}f_{ac} n)-\frac{1}{N^2}{\bigg(\sum_{n=n_0}^{n=n_1}\mbox{cos}(2{\pi}f_{ac} n)\mbox{sin}(2{\pi}f_{ac} n)\bigg)}^2\nonumber\\
& & -\frac{1}{N^3} \sum_{n=n_0}^{n=n_1}\mbox{sin}^2(2{\pi}f_{ac} n) {\bigg(\sum_{n=n_0}^{n=n_1}\mbox{cos}(2{\pi}f_{ac} n)\bigg)}^2 -\frac{1}{N^3} \sum_{n=n_0}^{n=n_1}\mbox{cos}^2(2{\pi}f_{ac} n) {\bigg(\sum_{n=n_0}^{n=n_1}\mbox{sin}(2{\pi}f_{ac}n)\bigg)}^2 \nonumber\\
& & +\frac{2}{N^3}\sum_{n=n_0}^{n=n_1}\mbox{cos}(2{\pi}f_{ac} n)\mbox{sin}(2{\pi}f_{ac}n)\sum_{n=n_0}^{n=n_1}\mbox{cos}(2{\pi}f_{ac}n)\sum_{n=n_0}^{n=n_1}\mbox{sin}(2{\pi}f_{ac}n).
\end{eqnarray}

The mean flow velocity $\bar{v}$ may then be written as

\begin{eqnarray}
\label{eqapp1_5}
v_c & = & \frac{1}{N^3 D}
\bigg(
\sum_{n=n_0}^{n=n_1}\mbox{sin}(2{\pi}f_{ac}n)
\sum_{n=n_0}^{n=n_1}\mbox{cos}(2{\pi}f_{ac} n)\mbox{sin}(2{\pi}f_{ac} n)-
\sum_{n=n_0}^{n=n_1}\mbox{cos}(2{\pi}f_{ac}n)
\sum_{n=n_0}^{n=n_1}\mbox{sin}^2(2{\pi}f_{ac} n)
\bigg)
\sum_{n=n_0}^{n=n_1} u_n \mbox{cos}(2{\pi}f_{ac}n)
\nonumber \\
& & + \frac{1}{N^3 D}
\bigg(
\sum_{n=n_0}^{n=n_1}\mbox{cos}(2{\pi}f_{ac}n)
\sum_{n=n_0}^{n=n_1}\mbox{cos}(2{\pi}f_{ac} n)\mbox{sin}(2{\pi}f_{ac} n)-
\sum_{n=n_0}^{n=n_1}\mbox{sin}(2{\pi}f_{ac}n)
\sum_{n=n_0}^{n=n_1}\mbox{cos}^2(2{\pi}f_{ac} n)
\bigg)
\sum_{n=n_0}^{n=n_1} u_n \mbox{sin}(2{\pi}f_{ac}n) \nonumber \\
& & +
\frac{1}{N^3 D}
\bigg(
\sum_{n=n_0}^{n=n_1}\mbox{cos}^2 (2{\pi}f_{ac} n)\sum_{n=n_0}^{n=n_1}\mbox{sin}^2(2{\pi}f_{ac} n)-{\bigg(\sum_{n=n_0}^{n=n_1}\mbox{cos}(2{\pi}f_{ac} n)\mbox{sin}(2{\pi}f_{ac} n)\bigg)}^2
\bigg)
\sum_{n=n_0}^{n=n_1} u_n.
\end{eqnarray}

Similarly, the acoustic parameters express as

\begin{eqnarray}
\label{eqapp1_6}
a_{ac} & = & \frac{1}{N^3 D}
\bigg(
\sum_{n=n_0}^{n=n_1}\mbox{sin}^2(2{\pi}f_{ac}n)
{\bigg(\sum_{n=n_0}^{n=n_1}\mbox{sin}(2{\pi}f_{ac} n)\bigg)}^2
\bigg)
\sum_{n=n_0}^{n=n_1} u_n \mbox{cos}(2{\pi}f_{ac}n) \nonumber \\
& & + \frac{1}{N^3 D}
\bigg(
\sum_{n=n_0}^{n=n_1}\mbox{cos}(2{\pi}f_{ac}n)
\sum_{n=n_0}^{n=n_1}\mbox{sin}(2{\pi}f_{ac}n)
-
\sum_{n=n_0}^{n=n_1}\mbox{cos}(2{\pi}f_{ac}n)\mbox{sin}(2{\pi}f_{ac}n)
\bigg)
\sum_{n=n_0}^{n=n_1} u_n \mbox{sin}(2{\pi}f_{ac}n) \nonumber \\
& & +\frac{1}{N^3 D}
\bigg(
\sum_{n=n_0}^{n=n_1}\mbox{sin}(2{\pi}f_{ac}n)
\sum_{n=n_0}^{n=n_1}\mbox{cos}(2{\pi}f_{ac} n)\mbox{sin}(2{\pi}f_{ac} n)-
\sum_{n=n_0}^{n=n_1}\mbox{cos}(2{\pi}f_{ac}n)\mbox{sin}^2(2{\pi}f_{ac} n)
\bigg)
\sum_{n=n_0}^{n=n_1} u_n,
\end{eqnarray}
and
\begin{eqnarray}
\label{eqapp1_7}
b_{ac} & = & \frac{1}{N^3 D}
\bigg(
\sum_{n=n_0}^{n=n_1}\mbox{cos}^2(2{\pi}f_{ac}n)
{\bigg(\sum_{n=n_0}^{n=n_1}\mbox{cos}(2{\pi}f_{ac} n)\bigg)}^2
\bigg)
\sum_{n=n_0}^{n=n_1} u_n \mbox{sin}(2{\pi}f_{ac}n) \nonumber \\
& & + \frac{1}{N^3 D}
\bigg(
\sum_{n=n_0}^{n=n_1}\mbox{cos}(2{\pi}f_{ac}n)
\sum_{n=n_0}^{n=n_1}\mbox{sin}(2{\pi}f_{ac}n)
-
\sum_{n=n_0}^{n=n_1}\mbox{cos}(2{\pi}f_{ac}n)\mbox{sin}(2{\pi}f_{ac}n)
\bigg)
\sum_{n=n_0}^{n=n_1} u_n \mbox{cos}(2{\pi}f_{ac}n) \nonumber \\
& & +\frac{1}{N^3 D}
\bigg(
\sum_{n=n_0}^{n=n_1}\mbox{cos}(2{\pi}f_{ac}n)
\sum_{n=n_0}^{n=n_1}\mbox{cos}(2{\pi}f_{ac} n)\mbox{sin}(2{\pi}f_{ac} n)-
\sum_{n=n_0}^{n=n_1}\mbox{sin}(2{\pi}f_{ac}n)\mbox{cos}^2(2{\pi}f_{ac} n)
\bigg)
\sum_{n=n_0}^{n=n_1} u_n.
\end{eqnarray}

\section{Asymtotic CRB}
\label{appendix2}

In this Appendix, we write the CRB of $v_c$ (\ref{eqdp8}), $V_{ac}$ (\ref{eqdp11}) and $\phi_{ac}$ (\ref{eqdp12}) respectively in both asymptotic cases 

\begin{equation}
\label{eqapp2_1}
2\gamma N \ll 1,
\end{equation}
and
\begin{equation}
\label{eqapp2_2}
2\gamma N \gg 1,
\end{equation}
where $\gamma$ and $N \equiv N_q$ are given by (\ref{eqdp7bis}) and (\ref{eq:nq}). Using (\ref{eq:nq}) and (\ref{eqdp7bis}), we note that (\ref{eqapp2_1}) and (\ref{eqapp2_2}) are respectively equivalent to

\begin{equation}
\label{eqapp2_1bis}
2\sqrt{2}\pi D_x F_{ac} \ll v_c,
\end{equation}
and
\begin{equation}
\label{eqapp2_2bis}
2\sqrt{2}\pi D_x F_{ac} \gg v_c.
\end{equation}

Firstly, we suppose that $2\gamma N \ll 1$ and that $\gamma \ll 1$ which means that the actual velocity signal corresponds to largely less than one acoustic period. The Taylor expansion at the 7th order of the sine functions in (\ref{eqdp8}), (\ref{eqdp11}) and (\ref{eqdp12}) respectively yields

\begin{equation}
\label{eqapp2_3}
\mbox{var}(v_c) \geq \sigma^2 \frac{45}{\pi^4 f_{ac}^4}\frac{1}{N^5},
\end{equation}
\begin{equation}
\label{eqapp2_4}
\mbox{var}(V_{ac}) \geq \frac{\sigma^2}{2} \frac{45}{\pi^4 f_{ac}^4}\frac{1}{N^5},
\end{equation}
\begin{equation}
\label{eqapp2_5}
\mbox{var}(\phi_{ac}) \geq \frac{\sigma^2}{2 V_{ac}^2} \frac{45}{\pi^4 f_{ac}^4}\frac{1}{N^5}.
\end{equation}

Using (\ref{eq:nq}) and (\ref{eqdp7bis}), we note that (\ref{eqapp2_3}-\ref{eqapp2_5}) may respectively be written as

\begin{equation}
\label{eqapp3_3bis}
\mbox{var}(v_c) \geq \sigma^2 \frac{45}{\pi^4 2^{5/2}}\frac{1}{D_x^5 F_e}\frac{v_c^5}{F_{ac}^4},
\end{equation}
\begin{equation}
\label{eqapp3_4bis}
\mbox{var}(V_{ac}) \geq \sigma^2 \frac{45}{\pi^4 2^{7/2}}\frac{1}{D_x^5 F_e}\frac{v_c^5}{F_{ac}^4},
\end{equation}
\begin{equation}
\label{eqapp3_5bis}
\mbox{var}(\phi_{ac}) \geq \sigma^2 \frac{45}{\pi^4 2^{7/2}}\frac{1}{D_x^5 F_e}\frac{v_c^5}{F_{ac}^4V_{ac}^2}.
\end{equation}

Writing (\ref{eq:sigma}) into (\ref{eqapp3_3bis}-\ref{eqapp3_5bis}) leads to

\begin{equation}
\label{eqapp2_3ter}
\mbox{var}(v_c) \geq \frac{1}{\mbox{SNR}} \frac{45}{\pi^4 2^{7/2}}\frac{1}{D_x^5 F_e}\frac{v_c^5 V_{ac}^2}{F_{ac}^4},
\end{equation}
\begin{equation}
\label{eqapp2_4ter}
\mbox{var}(V_{ac}) \geq \frac{1}{\mbox{SNR}}  \frac{45}{\pi^4 2^{9/2}}\frac{1}{D_x^5 F_e}\frac{v_c^5 V_{ac}^2}{F_{ac}^4},
\end{equation}
\begin{equation}
\label{eqapp2_5ter}
\mbox{var}(\phi_{ac}) \geq \frac{1}{\mbox{SNR}}  \frac{45}{\pi^4 2^{9/2}}\frac{1}{D_x^5 F_e}\frac{v_c^5}{F_{ac}^4},
\end{equation}
where $\mbox{SNR}$ is the linear signal-to-noise ratio. 

Secondly, we now suppose that $2\gamma N \gg 1$ which means that the actual velocity signal corresponds to largely great than one acoustic period. The asymptotic CRB is then such that

\begin{equation}
\label{eqapp2_6}
\mbox{var}(v_c) \geq \frac{\sigma^2}{N},
\end{equation}
\begin{equation}
\label{eqapp2_7}
\mbox{var}(V_{ac}) \geq \frac{2\sigma^2}{N}
\end{equation}
and
\begin{equation}
\label{eqapp2_8}
\mbox{var}(\phi_{ac}) \geq \frac{2\sigma^2}{N V_{ac}^2}.
\end{equation}

Using (\ref{eq:nq}), we note that (\ref{eqapp2_6}-\ref{eqapp2_8}) may respectively be written as

\begin{equation}
\label{eqapp2_6bis}
\mbox{var}(v_c) \geq \frac{\sigma^2}{\sqrt{2} D_x F_e} v_c,
\end{equation}
\begin{equation}
\label{eqapp2_7bis}
\mbox{var}(V_{ac}) \geq \frac{\sqrt{2} \sigma^2}{D_x F_e} v_c
\end{equation}
and
\begin{equation}
\label{eqapp2_8bis}
\mbox{var}(\phi_{ac}) \geq \frac{\sqrt{2} \sigma^2}{D_x F_e}\frac{v_c}{V_{ac}^2}.
\end{equation}

Lastly, inserting (\ref{eq:sigma}) into (\ref{eqapp2_6bis}-\ref{eqapp2_8bis}) finally leads to

\begin{equation}
\label{eqapp3_6bis}
\mbox{var}(v_c) \geq \frac{1}{\mbox{SNR}} \frac{1}{2^{3/2} D_x F_e} v_c V_{ac}^2,
\end{equation}
\begin{equation}
\label{eqapp3_7bis}
\mbox{var}(V_{ac}) \geq \frac{1}{\mbox{SNR}} \frac{1}{\sqrt{2} D_x F_e} v_c V_{ac}^2
\end{equation}
and
\begin{equation}
\label{eqapp3_8bis}
\mbox{var}(\phi_{ac}) \geq \frac{1}{\mbox{SNR}} \frac{1}{\sqrt{2} D_x F_e} v_c
\end{equation}

\end{document}